\documentclass{aa}
\input psfig.tex
\usepackage{graphicx}
\usepackage{natbib}

\usepackage{array}
\usepackage{graphics}
\usepackage{latexsym}
\usepackage{amssymb}
\usepackage{amsmath}
\usepackage{fancyhdr}
\usepackage{morefloats}
\usepackage{bm}
\bibpunct{(}{)}{;}{a}{}{,}

\begin{document}

\title{Theory of stellar population synthesis}
\subtitle{with an application to N-body simulations}

\author{S. Pasetto\inst{1},
			 	C. Chiosi\inst{2},
			  D. Kawata\inst{1}}

\offprints{sp2@mssl.ucl.ac.uk}

\institute{
University College London, Department of Space \& Climate Physics, Mullard Space Science Laboratory, Holmbury St. Mary, Dorking Surrey RH5 6NT, United Kingdom
\and
Department of Physics and Astronomy "Galileo Galilei", University of Padova, Padova, Italy }
\date{Accepted for pubblication in A\&A}

\titlerunning{Stellar populations and synthetic CMD}
\authorrunning{S.\ Pasetto, C.\ Chiosi, \& D.\ Kawata }

\abstract {}
{We present here a new theoretical approach to population synthesis. The aim is to predict colour magnitude diagrams (CMDs) for  huge numbers of stars. With this method we generate synthetic CMDs for N-body simulations of galaxies.
Sophisticated hydrodynamic N-body models of galaxies require equal quality simulations of the photometric properties of their stellar content. The only prerequisite for the method  to work is very little information on the star formation and chemical enrichment histories, i.e. the age and metallicity of all star-particles as a function of time.  The method  takes into account the gap between the mass of real stars and that of  the  star-particles in N-body simulations, which best correspond to the mass of star clusters with different age and metallicity, i.e. a manifold of single stellar sopulations (SSP). } 
{The theory extends the concept of SSP  to include the phase-space (position and velocity) of each star. Furthermore,  it accelerates the building up of simulated  CMD by using a  database of theoretical SSPs that extends to all ages and metallicities of interest. Finally, it uses the concept of distribution functions to build up the CMD. The technique is independent of the mass resolution and the way the N-body simulation has been calculated. This allows us to generate CMDs for simulated stellar systems of any kind: from open clusters to globular clusters, dwarf galaxies, or spiral and elliptical galaxies.  }
{The new theory is applied to an N-body simulation of a disc galaxy to test its performance and highlight its flexibility.} 
{}


\maketitle

\section{Introduction}\label{Introduction}
In the past few decades the unprecedented development of the computational facilities drastically increased the number of NB-based astrophysical simulations. In some areas of astrophysics, this powerful approach represents  the only viable laboratory experiment because gravitational interaction on an astrophysical scale  is clearly impossible to reproduce in any laboratory.

Since the pioneering works of \citet[][]{1960ZA.....50..184V} and \citet[][]{1963MNRAS.126..223A}, the orbit integration techniques for the NB problem (with N the number of gravitationally interacting bodies) greatly improved \citep[see, e.g.,][]{1988csup.book.....H,2003gnbs.book.....A} both theoretically, with the developments of the tree-codes, particle-mesh (PM), particle-particle-particle mesh (P$^3$M) or see, e.g., \citet[][]{2011EPJP..126...55D} for a recent review, and technically, by exploiting the message-passing-interface protocol for the use of massively parallel machines.
Although that the N-body simulations were originally conceived as a tool to follow the integration of the orbits, the importance of including energy dissipative processes soon became evident. The treatment of baryonic interactions in N-body simulations is nowadays standardised by popular protocols such as  smoothed particle hydrodynamics (SPH) or adaptive-mesh-refinement\footnote{Although AMR codes are not an NB-type code, they often use star-particles to describe the stellar components. Therefore, we here include AMR for our terminology of N-body simulations in this paper.} (AMR) implemented in several codes, e.g.,  Algodoo \citep[e.g.,][]{2012PhTea..50..278K}, DualSPHysics \citep[e.g.,][]{GomezGesteira2012}, SPH-flow \citep[e.g.,][]{Oger2006803}, ENZO \citep[e.g.,][]{2004astro.ph..3044O}, EvoL \citep[e.g.,][]{2010A&A...513A..36M}, GCD+ \citep[e.g.,][]{2003MNRAS.340..908K}, GADGET \citep[e.g.,][]{2001NewA....6...79S}, Gasoline \citep[e.g.,][]{Wadsley2004137}, FLASH \citet[e.g.,][]{Dubey2008a} or RAMSES \citep[e.g.,][]{2010ascl.soft11007T,2012arXiv1202.6400F}, which also includes (in astrophysical context) a sticky-particle approach \citep[][]{2002A&A...392...83B, 2010ApJ...714L.275M}. Thus, even though the recipes for implementing the physics of dissipative phenomena are still controversial, several NB codes are currently able to evolve with time the baryonic component, i.e. gas and stars, in their mutual interaction.
Therefore, it is nowadays possible and mandatory to develop the right tools to compare the results of dynamical N-body simulations with observational data for the stellar content of galaxies.

\citet[][]{2010A&A...518A..43T} recently developed a technique to derive the integrated spectra,  magnitudes and colours of the stellar content  of simulated galaxies. These authors focused on the evolution of non-resolved stellar populations and exploited the concept of the spectral energy distribution (SED) of  an SSP in the context of discrete resolved-mass points of an N-body simulation. There, the integrated monochromatic flux in a  given photometric system (set of discrete pass-bands $\Delta\lambda $) of a galactic stellar component at the time $t$  and metallicity $Z$, ${F_\lambda }\left( {Z;t} \right)$, is taken to be the convolution of the star formation rate (SFR)  $\psi \left( {Z;t} \right)$  (which is assumed to depend only on age and metal content), and the integrated monochromatic flux of constituent SSPs, ${\phi _\lambda }\left( {Z;t} \right)$, i.e. ${F_\lambda }\left( {Z;t} \right) = \psi \left( {Z;t} \right) * {\phi _\lambda }\left( {Z;t} \right)$ \footnote{Where $f * g$ is the standard convolution operator between the generic functions $f$ and $g$. In general, we can indeed always write ${F_\lambda } = {L^{ - 1}}\left[ {L\left[ \psi  \right]L\left[ {{\phi _\lambda }} \right]} \right]$, with $L\left[ \psi  \right]$ the discrete Laplace transform of the SFR we obtain directly from the N-body simulations and $L\left[ {{\phi _\lambda }} \right]$ the Laplace transform of the monochromatic integrated flux of the a simple stellar population.}.

In this study we present a new  theory of population synthesis that is particularly designed to manage very many stars, and is based on the concept of a distribution function (DF). We focus then on generating the CMDs for the stellar system which are simulated with advanced hydro-dynamical N-body simulation techniques. The subject is particularly timely in view of the modern capabilities of data acquisition of star-by-star photometry in nearby galaxies. This is thanks to modern instrumentation already in place or in progress for the coming years, and also in view of the N-body simulations of model galaxies, in which sampling of either real or fake stars with many millions of objects is possible.

The only prerequisite for the method  to work is that minimal information about the history of star formation and chemical enrichment is implemented in and read off the  N-body simulations. For each star-particle of the N-body simulation we must know the age at which it was born and the chemical composition (metallicity) of the parental medium.
The new formulation of population synthesis takes naturally into account  the gap  between the mass of the real stars and the mass of the star-particles in N-body simulations if these latter are closer to the mass of a star cluster than the mass of a real star. Therefore the star-content of an N-body simulation is better described as a manifold of star clusters with different age and metallicity, i.e. a manifold of SSP whose photometric properties, primarily the CMD, are well known. Second, CMDs containing an arbitrary number of stars can be easily simulated by defining and making use of the concept of DF of  stars in the CMD.

This novel technique is applicable to the stellar content of single open/globular clusters, to our own Galaxy, to that of   nearby galaxies within the Local Group, and to all galaxies of the local Universe whose stellar content can be resolved into stars.  Moreover, the algorithm can be easily interfaced with other codes, e.g.  Galaxia \citep[e.g.,][]{2011ApJ...730....3S}, or  the Galaxy Simulators  HRD-GST \citep[e.g.,][]{1995A&A...295..655N,1996A&A...310..771N,1996A&A...315..116N,1997A&A...324...65N,2002A&A...392.1129N,2006A&A...451..125V}, and Besancon model \citep[e.g.,][]{2003A&A...409..523R}, thus  extending their performances  with the advantages that will become soon clear in  what follows.

Setting up the new  technique, we focused our attention on a few key requirements that should be met:
\begin{enumerate}
	\item The algorithm has to be independent of the particular N-body simulation it is applied to. Thus, it is designed to accept the output of a simulation as input, without interfering with the NB integration.

	\item The algorithm must not depend on the specific prescriptions used to generate the stellar models that are adopted to describe the composite stellar populations and associated CMDs.

	\item The technique needs to handle CMDs populated by several billions of stars, i.e. which are potentially representative of the largest systems of stars known (e.g., giant elliptical galaxies). Indeed, our algorithm should be applicable to every resolved stellar population no matter what the nature of the stellar system under consideration. Therefore CMDs for giant elliptical or spiral galaxies that contain up to ${10^{12}}$ stars have to be synthesized with agility.

	\item Finally, we require the code to only require little  computational resources. The typical CMD of point 3, has to be obtained on a serial machine, i.e. its realization has to avoid more complicated MPI/OpenMP programming protocols.
\end{enumerate}

To achieve those goals, we have developed a method for synthesising CMDs that takes all  the above requirements into account. The plan of the paper is as follows: in Section \ref{TotSP}  we  extend some of the theory concepts of the stellar populations within the framework of the DFs. In Section \ref{Sp4Nb} we particularise the general theory  to the case of the N-body simulations. In Section \ref{CMDrp} we deal with the simulations of CMDs that contains huge numbers of stars. In Section 5 we present an example of a synthetic CMD realization. 
 In Section \ref{NB_simul} we show an example of the technique.  Finally we draw some concluding remarks in Section \ref{ConcDisc} and  briefly mention some possible  applications of the new method to the synthesis of stellar populations.

\section{Theory of Stellar Populations}\label{TotSP}
We extend here the theory of stellar populations to systems (assemblies of stars) with an assigned distribution in the phase-space. This completes the standard theory of  stellar populations \citep[e.g.,][]{2005essp.book.....S, 2011spug.book.....G} by assigning a position and velocity to each star.

Every real (or realistically simulated) stellar system is a set of stars born at different times and positions, and with different velocities, masses and chemical compositions. We call this assembly of stars  \textit{composite-stellar-population} (CSP).
The position ${\mathbf{x}}$ of each star of a CSP  is a point  in the space $\mathbb{X}$ of coordinates (or space of configurations), and its linear momentum ${\mathbf{p}}$ is geometrically referred to as a fiber of the cotangent space ${\text{T}}_{\mathbf{x}}^*\mathbb{X}$ of $\mathbb{X}$ at ${\mathbf{x}}$ \citep[][]{1998cdca.book.....J}. Therefore, the  phase-space of a CSP is  the cotangent bundle ${{\text{T}}^*}\mathbb{X} \equiv \bigcup\limits_{\mathbf{x}} {{\text{T}}_{\mathbf{x}}^*\mathbb{X}} $ of its configuration space $\mathbb{X}$. It follows that  ${{\text{T}}^*}\mathbb{X}$ is a $\dim [{\text{T}}_{}^*\mathbb{X}] = 6N$  manifold,  with $N$ the total number of  stars in the CSP. Each  point in it defines the phase-space of our CSP. We indicate this manifold with the symbol ${\mathbf{\Gamma }}$, i.e.  ${\mathbf{\Gamma }} \equiv {{\text{T}}^*}\mathbb{X}$ where ${\mathbf{\Gamma }} = \left( {{\mathbf{x}},{\mathbf{v}}} \right) = \left( {{x_1},{x_2},...,{x_{3N}},{v_1},{v_2},...,{v_{3N}}} \right)$. Moreover, to completely define the CSP parameters at a generic time $t$, we need to specify the distribution in the space of the masses, $M$,  and metallicities, $Z$ of all its members. We  call this extended phase-space the \textit{existence-space} of the CSP, i.e.  $\mathbb{E} \equiv M \times Z \times {\mathbf{\Gamma }}$ of the CSP at the time $t$ and $\mathbb{E} \times \mathbb{R}$ the \textit{extended-existence-space with time} $t$. In $\mathbb{E} \times \mathbb{R}$, the stars evolve with time, i.e. they continuously move in space,   lose mass,  enrich in metals, and move in the phase-space. Hence we can safely define in $\mathbb{E}$ the distribution function (DF) for the CSPs under the assumption of continuity and differentiability. Let us consider  a sample of identical systems whose initial conditions span a certain volume of the space $\mathbb{E}$. We refer to this sample simply as an \textit{ensemble}, borrowing the name-root from the grand microcanonical ensemble adopted in Statistical Mechanics. The number of systems $dN$ at the time $t$ with mass within $dM$, metallicity within $dZ$ and phase-space within $d{\mathbf{\Gamma }} = \left( {d{\mathbf{x}},d{\mathbf{v}}} \right)$ is given by

\begin{equation}\label{Eq1}
	dN = N{f_{{\text{CSP}}}}\left( {M,Z,{\mathbf{\Gamma }};t} \right)dMdZd{\mathbf{\Gamma }},
\end{equation}

\noindent with ${f_{{\text{CSP}}}}\left( {M,Z,{\mathbf{\Gamma }};t} \right)$ the  DF in $\mathbb{E}$, ${f_{{\text{CSP}}}}:{\mathbb{R}^ + } \times {\mathbb{R}^ + } \times {\mathbb{R}^{6N}} \times \mathbb{R} \to {\mathbb{R}^ + } $ continuous and with partial derivative continuous (where $\mathbb{R}$ is the set of all real numbers, and ${\mathbb{R}^ + } = \left\{ {a|a \in \mathbb{R} \wedge a \geqslant 0} \right\}$) and  the total number of systems in the \textit{ensemble} is fixed by normalising to one the DF in the space $\mathbb{E}$, i.e.

\begin{equation}\label{Eq2}
	\int_{}^{} {{f_{{\text{CSP}}}}\left( {M,Z,{\mathbf{\Gamma }};t} \right)dMdZd{\mathbf{\Gamma }} = 1}.
\end{equation}

\noindent We proceed to formally \textit{define} an SSP as follows. At a given time $t$, we divide $\mathbb{E}$ into a grid in terms of discrete intervals of metallicity $dZ$ and phase-space $d\Gamma $. Then every infinitesimal unit (i.e. every elementary cell) of this sub-space of $\mathbb{E}$ defines an SSP.

We assume that every CSP originates at time ${t_0}$ from an episode of single star formation  in a chemically homogeneous medium. The stars of the CSP have  the mass spectrum ${f_M} = {f_M}\left( {M,{t_0}} \right) \propto \xi \left( {M,{t_0}} \right)$ where $\xi \left( {M,{t_0}} \right)$ is the initial mass function (IMF). Their   metallicity distribution is ${f_Z} = {f_Z}\left( {{t_0}} \right) \propto {Z_{{\text{CSP}}}}\left( {{t_0}} \right)$. Finally, their  phase-space distribution function is ${f_{\mathbf{\Gamma }}} = {f_{\mathbf{\Gamma }}}\left( {{\mathbf{\Gamma }};{t_0}} \right)$. The DF of the CSP can be written as
\begin{equation}\label{Eq3}
	{f_{{\text{CSP}}}} = \sum\limits_{i = 0}^n {{f_{{\text{SSP}}}}},
\end{equation}

\noindent where $n$ is the number of SSP, with eventually $n \to \infty $, and ${f_{{\text{SSP}}}} = {f_{{\text{SSP}}}}\left( {M,{Z_0},{{\mathbf{\Gamma }}_0};{t_0}} \right)$ is the DF of SSP  of which  $Z = {Z_0}$, ${\mathbf{\Gamma }} = {{\mathbf{\Gamma }}_0}$, $t = {t_0}$ are the metallicity, SSP's phase-space and age  at the instant ${t_0}$, respectively.

As time elapses, this stellar population ages. According to their mass,  the stars eventually leave the main-sequence (MS) after a time ${t_{{\text{MS}}}}=t_{{\text{MS}}}(M)$ and soon after die (supernovae)  or enter a quiescence stage (white dwarfs), injecting metals into the interstellar medium in form of supernova explosions or quiet winds.  The interstellar medium becomes richer in metals due both to self-enrichment by  the SSP under consideration  and the contribution from all other stellar SSPs.
At the generic time $t > {t_0}$, i.e. after an elapsed time $\tau  \equiv t - {t_0}$  otherwise known as  the age of the stellar population, the SSP has processed stellar material. As a result of this activity, an age-metallicity relationship   ${Z_{{\text{CSP}}}} = {Z_{{\text{CSP}}}}\left( t \right)$ is built up for the CSP.

In addition to this, the number of trajectories in the phase-space entering a volume $d{\mathbf{\Gamma }}$, will in general be different from the number of those leaving the same volume. Thus  ${f_{\mathbf{\Gamma }}}$ evolves following Liouville's equation of the form

\begin{equation}\label{Eq4}
	\frac{{\partial {f_{\mathbf{\Gamma }}}}}{{\partial t}} =  - \iota \mathcal{L}{f_{\mathbf{\Gamma }}} =  - \left( {\frac{\partial }{{\partial {\mathbf{\Gamma }}}} \cdot {\mathbf{\dot \Gamma }} + {\mathbf{\dot \Gamma }} \cdot \frac{\partial }{{\partial {\mathbf{\Gamma }}}}} \right){f_{\mathbf{\Gamma }}},
\end{equation}
which is independent of the nature of the equation of motion (e.g., its correctness in the form of Eq. \eqref{Eq4} does \textit{not} require the existence of a Hamiltonian). $\mathcal{L}$ is the Liouvillean operator, eventually extended to account for a creation function $C\left( {{f_{{\text{CSP}}}};t} \right)$, and $\iota$ the imaginary unit. The formal solution of this equation is given by the Taylor expansion series of the time dependence of ${{f_{\mathbf{\Gamma }}}\left( {{\mathbf{\Gamma }};{t}} \right)}$ around ${{f_{\mathbf{\Gamma }}}\left( {{\mathbf{\Gamma }};{t_0}} \right)}$
\begin{equation}\label{Eq5}
	{f_{\mathbf{\Gamma }}} = {e^{-\iota \mathcal{L}t}}{f_{\mathbf{\Gamma }}}\left( {{\mathbf{\Gamma }};{t_0}} \right)=\sum\limits_{m = 0}^\infty  {\frac{{{{\left( { - t} \right)}^m}}}{{m!}}\frac{{{\partial ^m}}}{{\partial {t^m}}}{f_{\mathbf{\Gamma }}}\left( {{\mathbf{\Gamma }};{t_0}} \right)}, 
\end{equation}
and ${e^{ - \iota \mathcal{L}t}} = \sum\limits_{m = 0}^\infty  {\frac{{{{\left( { - t} \right)}^m}}}{{m!}}{{\left( {\iota \mathcal{L}} \right)}^m}}$ defines the infinite series of operators applied on any function on its right side.
In this way, the initial SSP has generated a CSP whose distribution in the existence space of the CSP, $\mathbb{E}$, has a DF  ${f_{{\text{CSP}}}}\left( {M,Z,{\mathbf{\Gamma }};t} \right)$. \\
Now that we have stated these fundamental concepts, it is easy to proceed with the following formal definitions to recover the usual concepts of the classical synthesis of population theory:

\begin{itemize}

\item \textsf{Present-day-mass-function:} The integral of the DF over the metallicity $Z$ and phase-space $\Gamma$ (from Eq. \eqref{Eq1})
\begin{equation}\label{Eq6}
	\int_{}^{} {N{f_{{\text{CSP}}}}\left( {M,Z,{\mathbf{\Gamma }};t} \right)dZd{\mathbf{\Gamma }}}  = \hat \xi \left( {M;t} \right),
\end{equation}
yields the ``present''-day-mass-function  $\hat \xi \left( {M;t} \right)$  (PDMF), i.e. the total number of stars per mass interval at the time $t$. This can be expressed  by the approximate relation

\begin{equation}\label{Eq7}
	\hat \xi \left( {M;t} \right) = \left\{ {\begin{array}{*{20}{c}}
	  {\xi \left( M \right)\frac{{{t_{{\text{MS}}}}}}{t-t_0}}&{{t_{{\text{MS}}}} < \tau} \\
	  {\xi \left( M \right)}&{{t_{{\text{MS}}}} > \tau,}
	\end{array}} \right.
\end{equation}
where $\xi \left( M \right)$ is IMF of the MS stars. More fine-tuned approaches based on the evolutionary flux, total luminosity and fuel-consumption theorem can be easily worked out upon necessity.\\

\item  \textsf{Age-metallicity function}: Integrating the DF over the mass $M$ and phase-space $\mathbf{\Gamma }$
\begin{equation}\label{Eq8}
	\int_{}^{} {N{f_{{\text{CSP}}}}\left( {M,Z,{\mathbf{\Gamma }};t} \right)d{\mathbf{\Gamma }}dM}  = \chi \left( {Z,t} \right),
\end{equation}
we obtain the age-metallicity relation, i.e. the number of stars formed per metallicity interval at the time $t$. \\

\item \textsf{The phase-space distribution-function}: Integrating upon the mass $M$ and metallicity $Z$
\begin{equation}\label{Eq9}
	\int_{}^{} {M{f_{{\text{CSP}}}}\left( {M,Z,{\mathbf{\Gamma }};t} \right)dMdZ}  = {e^{ - \iota \mathcal{L}  t }}{f_{\mathbf{\Gamma }}}\left( {{\mathbf{\Gamma }};{t_0}} \right),
\end{equation}
where we made use of  Eq. \eqref{Eq5}, yields the the phase-space DF.
\end{itemize}

In a natural way, we can extend the approach used in defining Eqs. \eqref{Eq6}, \eqref{Eq8} and \eqref{Eq9} by performing the following monodimensional integration:

\textsf{Metallicity phase-space relationship}: We consider the following integration
\begin{equation}\label{Eq10}
	\eta \left( {Z,{\mathbf{\Gamma }};t} \right) \equiv  \int_{}^{} {{f_{{\text{CSP}}}}\left( {M,Z,{\mathbf{\Gamma }};t} \right)dM}
\end{equation}
to define the metallicity-phase-space relationship. Very often the projection of this function onto the configuration space is used
\begin{equation}
\int_{}^{} {\eta\left( {Z,{\mathbf{\Gamma }};t } \right){d^{3N}}{\mathbf{v}}}  = {\hat \eta}\left( {Z,{\mathbf{x}};t } \right),
\end{equation}
where ${d^{3N}}{\mathbf{v}}$ is the 3N velocity element of the ${\mathbf{\Gamma }}$ space. This plays an important role, e.g., in relation to  radial metallicity gradients of resolved populations in external galaxies and the vertical metallicity gradients in the Milky Way, i.e. ${\hat \eta}\left( {Z,{\mathbf{x}};t } \right)=0$ $\forall t$ can implicitly define a function that relates the metallicity $Z$ and the radius of a galaxy $R$ once cylindrical coordinates are in use.

\textsf{Mass-phase-space relationship}: In the same way we define
\begin{equation}\label{Eq11}
\mu\left( {M,{\mathbf{\Gamma }};t} \right) \equiv 	\int_{}^{} {{f_{{\text{CSP}}}}\left( {M,Z,{\mathbf{\Gamma }};t} \right)dZ },
\end{equation}
i.e. the mass-phase-space relation.  More often its projection onto the configuration space

\begin{equation}
\int_{}^{} {\mu\left( {Z,{\mathbf{\Gamma }};t } \right){d^{3N}}{\mathbf{v}}}  = {\hat \mu}\left( {Z,{\mathbf{x}};t } \right)
\end{equation}
is used, which is related, e.g., to the evolution of the globular clusters and the mass-segregation effects, i.e. we can use it to express the higher concentration of massive (or binary stars) through the centre of the cluster with respect to the less massive stars \citep[e.g.,][]{1987degc.book.....S}.

\textsf{Mass-Metallicity relationship}: Finally, we define
\begin{equation}\label{Eq12}
	\varpi(M,Z;t) \equiv \int_{}^{} {{f_{{\text{CSP}}}}\left( {M,Z,{\mathbf{\Gamma }};t} \right)d{\mathbf{\Gamma }} },
\end{equation}
i.e. the mass-metallicity relation for a resolved stellar population.  This relation  is of paramount importance e.g., when the masses of the stars are obtained by parallaxes, transits,  and astero-seismology.  For nearby galaxies  this relation is  relevant  only in studies of integrated photometry.

\section{Stellar populations in  N-body simulations}\label{Sp4Nb}
The concepts we have just  developed can be straightforwardly applied  to  N-body simulations of any kind, from those for globular clusters with single or multiple stellar populations \citep[e.g.,][]{2012ApJ...744...58M,2004ApJ...612L..25N}, to those for the  Milky Way stellar fields  observed along any line of sight  \citep[e.g.,][]{1995A&A...295..655N,2006A&A...451..125V}, to the dwarf galaxies of the Local Group \citep[e.g.,][]{1997RvMA...10...29G,1998ARA&A..36..435M,2009ARA&A..47..371T}, or even  external galaxies once they are resolved into individual stars \citep[e.g.,][]{2002AJ....123.3108H,2011A&A...526A.123R,2011A&A...530A..58C}. Indeed, in the adopted formulation,  all the dynamical aspects of the processes governing the formation of the astronomical object under investigation are considered through the most general Liouvillean operator, whose form can be specified case by case.  In this context, it is worth recalling that  a mutual dependence of star formation, which is responsible for the building up of the stellar population of a stellar system,  and the dynamical processes that  govern the large scale aggregation of it, is in general supposed to occur  in extended many-body interactions  \citep[e.g.][]{2011A&A...525A..99P,2012A&A...542A..17P}. \\

\textsf{Star-particles in N-body simulations}: Before proceeding further, we must suitably link the definition and properties of CSPs to the elemental building blocks of N-body simulations, i.e. the  ``star-particles''.

We \textit{define }an N-body simulation as a stochastic realization of ${f_{{\text{CSP}}}}\left( {M,Z,{\mathbf{\Gamma }};\tau } \right)$ in the existence space $\mathbb{E}$ of the CSPs. We focus our attention on the evolution of a single star-particle  in an N-body simulation. This particle $p$ is born at the instant ${t_{0,p} \equiv t_p}$ with metallicity ${Z_p} = {Z_p}\left( {{t_p}} \right)$, at a position in the single particle phase-space ${\mathbf{\gamma }}$ (a sub-manifold of ${{\text{T}}^*}\mathbb{X}$) given by ${{\mathbf{\gamma }}_p} = {{\mathbf{\gamma }}_p}\left( {{t_p}} \right)$, with a certain metallicity $Z=Z(t_p)$ (the metal content of the parent gas component at time $t_p$),  and with a  mass ${M_p} = {M_p}\left( {{t_p}}\right)$.

The mass of the star-particles and the smaller mass resolution $\tilde{M}$ of the N-body simulations is one of the greatest problems of modern computational N-body simulations and companion mathematical techniques. Apart from a few recent N-body simulations that were specifically tailored for star clusters \citep[e.g.,][]{2003NewA....8..337H,2011MNRAS.411.1989Z,2012MNRAS.420.1503P}, the particle mass of the modern N-body simulations dedicated galaxies and galaxy clusters can vary within a wide range,  typically ${\log _{10}}\left( {{\tilde{M}}} \right) \simeq 2 \div 8$ \citep[e.g.,][]{2009PASJ...61..481S,2008PASJ...60..667S,2011ApJ...742...76G,2009MNRAS.399L.174O}. 
This means that the average mass of the star-particles in the NB simulations is larger than the average mass of the stars in a SSP. Therefore each star-particle is  representative (in mass) of many hundreds stars (or more). However, since all stars in a star-particle of a simulation are assumed to be born at the same time with the same metallicity, the stellar content of the star-particle itself is well described by an SSP (see SSP definition in Section \ref{TotSP}). Nevertheless, NB simulations cannot easily describe the phase-space distribution of individual stars within a star-particle. In this paper we assume that all stars within each star-particle share the same position in the phase-space.
As a consequence of this, the volume of the existence space $\mathbb{E}$ occupied by elemental cells is  bounded by the mass resolution, say $d{\mathbb{E}_{\min }} \equiv d{M_p} \times dZ \times d{\mathbf{\Gamma }}$, and it is generally larger than the volume necessary to properly map a CSP, $d\mathbb{E} \equiv d\tilde{M} \times dZ \times d{\mathbf{\Gamma }}$.

On  one hand,  if the problem clearly shows that  attempts to map the complete IMF are at present out of reach, this indicates on the other hand how a possible solution has to be searched for in the concept of SSP itself.

In an N-body simulation, at the time $t > {t_p}$  a star-particle is located in ${{\mathbf{\gamma }}_p} = {{\mathbf{\gamma }}_p}\left( t \right)$,  has metallicity ${Z_p} = {Z_p}\left( t \right)$ and mass ${M_p} = {M_p}\left( t \right)$. Therefore,  mass, metallicity and position in the single-particle phase-space univocally identify  the NB-particle all along its evolutionary history  from ${t_{p}=t_{0,p}}$ to  $t$. At the initial time ${t_p}$ for each particle $p$ we can associate an SSP to the star-particle in a natural way, provided that some re-scaling of the SSP mass, ${M_{{\text{SSP}}}}$, is made in relation to the star-particle mass, ${M_p}$.   From Eq. \eqref{Eq2}, the total mass of the SSP can be written as follows:

\begin{equation}\label{Eq13}
	\begin{gathered}
	  \int_{}^{} {M N {f_{{\text{CSP}}}}\left( {M,Z,{\mathbf{\Gamma }};t} \right)\delta \left( {Z - {Z_p},{\mathbf{\Gamma }} - {{\mathbf{\Gamma }}_p},{t-t_p}} \right)dMdZd{\mathbf{\Gamma }}}    \hfill \\
=	  \int_{}^{} {M N {f_{{\text{CSP}}}}\left( {M,{Z_p},{{\mathbf{\Gamma }}_p};{t_p}} \right)dM}    \hfill \\
=	  \int_{}^{} {M\hat \xi \left( {M;{t_p}} \right)dM}    \hfill \\
=	  \int_{}^{} {M\xi \left( M \right)dM}  = {M_{{\text{SSP}}}}\left( {{t_p}} \right), \hfill \\
	\end{gathered}
\end{equation}
where $\delta $ is the ordinary multidimensional Dirac function, and in the last row we used Eq. \eqref{Eq7} with ${t_p} = t_{0,p} = 0 < {t_{{\text{MS}}}}  \forall M$. In other words, the SSP mass written in terms of the DF $f_{\text{CSP}}$ or the IMF.

In the same way, we can cast the DF ${f_{{\text{CSP}}}}\left( {M,Z,{\mathbf{\Gamma }};t} \right)$ of the CSP as a function of the star-particle mass ${M_p}$ and vice versa. The following identities can be written:
\begin{align}\label{Eq14}
  & \int {\frac{{{M_p}}}{{{M_{{\text{SSP}}}}}}M  N {f_{{\text{CSP}}}}\left( {M,{Z_p},{{\mathbf{\Gamma }}_p};{t_p}} \right)dM}  \nonumber\\
  & \equiv \int {M N {f_{{\text{CSP,}}p}}\left( {M,{Z_p},{{\mathbf{\Gamma }}_p};{t_p}} \right)dM}  \nonumber\\
  & = \int {M N {f_{{\text{SSP,}}p}}\left( {M,{Z_p},{{\mathbf{\Gamma }}_p};{t_p}} \right)dM}  \nonumber\\
  & = {M_p}\left( {{t_p}} \right).
\end{align}

\noindent While associating the SSP to a generic star-particle born at time $t_p$ with metallicity $Z_p(t_p)=Z_p$ and mass $M_p(t_p)=M_p$ is straightforward, at a generic time $t>t_p$ it is no longer so simple. But if the mass and metallicity of a star-particle do not change with time, the SSP associated to a generic star-particle  will simply  evolve and age, losing a certain amount of mass in form of gas, changing a fraction of its living stars into remnants, and changing its spectral and photometric properties in a way that is fairly well known from the theory of stellar evolution and population synthesis \citep[e.g.,][]{1991AJ....102..951T,1994A&AS..106..275B,1996ApJ...462..672T,1997AJ....114..680A,2001ApJS..136...25H,2005ARA&A..43..387G,2008A&A...484..815B,2009A&A...508..355B}.

\section{Huge CMDs at no cost}\label{CMDrp}
 With the formalism we developed on the basis of the CMD of  CSP, we now tackle the key question: how can we build CMDs for billions of stars in practice at less computational cost? Because real CPS are the result of the past history of star formation in clusters and fields at the complexity levels of galaxies, how can we easily achieve the same level of information in numerical simulations?

\subsection{Associating an SSP to a star-particle}
The SSP corresponding to the ${p^{th}}$ star-particle is obtained by two-dimensional interpolation (in age and metallicity). The number of stars $d{\nu _p}\left( {L,{T_{{\text{eff}}}}} \right)$ for the interval of luminosity $L$ and effective temperature ${T_{{\text{eff}}}}$, $dLd{T_{{\text{eff}}}}$ results from

\begin{equation}\label{Eq21}
	d{\nu _p}\left( {{T_{{\text{eff}}}},L} \right) = {f_{{\text{SSP}},p}}\left[\kern-0.15em\left[ {\frac{{d\left( {\tau ,Z} \right)}}{{d\left( {{T_{{\text{eff}}}},L} \right)}}}
 \right]\kern-0.15em\right]d{T_{{\text{eff}}}}dL,
\end{equation}
where we referred, by extension of the Jacobean-matrix formalism, with $\left[\kern-0.15em\left[ {\frac{{d\left( {\tau ,Z} \right)}}{{d\left( {{T_{{\text{eff}}}},L} \right)}}}
 \right]\kern-0.15em\right]$ to the transformation applied on two of the dimensions in which SSP is defined: from age-metallicity space to the  effective temperature - luminosity space. Or analogously,	

\begin{equation}\label{Eq22}
d{\nu _p}\left( {C,m} \right) = {f_{{\text{SSP}},p}}\left[\kern-0.15em\left[ {\frac{{d\left( {\tau ,Z} \right)}}{{d\left( {{T_{{\text{eff}}}},L} \right)}}}
 \right]\kern-0.15em\right]\left[\kern-0.15em\left[ {\frac{{d\left( {{T_{{\text{eff}}}},L} \right)}}{{d\left( {C,m} \right)}}}
 \right]\kern-0.15em\right]dCdm
 \end{equation}
for the same transformation to the space of the colour $C$ and magnitude $m$.
In general there is no analytic formulation for the two matrixes $\left[\kern-0.15em\left[ {\frac{{d\left( {\tau ,Z} \right)}}{{d\left( {{T_{{\text{eff}}}},L} \right)}}}
 \right]\kern-0.15em\right]$ and $\left[\kern-0.15em\left[ {\frac{{d\left( {{T_{{\text{eff}}}},L} \right)}}{{d\left( {C,m} \right)}}}
 \right]\kern-0.15em\right]$. They are derived numerically from the tabulations of bolometric corrections, and colours of, e.g., the Johnson-Cousin-Glass system. 

The key idea of Eq. \eqref{Eq21} or \eqref{Eq22} is to describe a CMD as a matrix, i.e. a projected DF, whose elements are the relative frequency (or percentage) of stars of different colour and magnitude  in some photometric system, per elemental area of the CMD. For a given photometric system with pass-bands $\Delta \lambda_\alpha$ (for instance $\alpha$ stands for U, B, V,...) for which magnitudes $m_\alpha$ and colours $C_{\alpha \beta}\equiv m_\alpha-m_\beta$ can be computed (magnitudes and colours can be either absolute and intrinsic or apparent and reddened\footnote{The same considerations also apply to bolometric luminosities and effective temperatures.}), Eq. \eqref{Eq22} represents a 2D grid of elemental square cells $d{\nu _p}\left( {{C^c_{\alpha \beta}},{m^c_\alpha}} \right)$, identified by the coordinates of their centres $m_\alpha^c$ and  $(m_\alpha-m_\beta)^c$. The path drawn in the CMD by a  single SSP of assigned age and chemical composition, see Section \ref{TotSP} and Eq. \eqref{Eq22}, will occupy a number of cells from  the main sequence  to  the last observable stage. Each cell is populated by a certain number of stars according to the underlying luminosity function of the SSP, which in turn is related to the evolutionary rate  and IMF (see Eq. \eqref{Eq6} and \eqref{Eq7}).  If many SSPs are present, as in the case of a composite CSP, each cell  contains a total number of stars given by the sum of all contributions by different SSPs passing through this cell. Using Eqs. \eqref{Eq3} together with \eqref{Eq22}, we obtain
\begin{equation}\label{nusomma}
	{\nu _p}\left( {{C_{\alpha \beta}},{m_\alpha}} \right) = \sum\limits_p^{} {d{\nu _p}\left( {{C_{\alpha \beta}},{m_\alpha}} \right)}
\end{equation}
 Therefore, a relative percentage of stars is associated to each cell (matrix element) with respect to the total.  The number of stars per cell can  be easily normalised according to the problem under investigation. The observational CMDs that contain an arbitrary number of stars, from a few thousands to billions, is reduced to a matrix of number frequencies (relative percentages, Eq. \eqref{Eq22}). The prescription can be easily extended to include any history of star formation of any intensity, $\psi $, as well as kinematical effects (${f_{\mathbf{\Gamma }}}$ as in Eq. \eqref{Eq5}).

\begin{figure*}
\centering
{\includegraphics[width=8cm, height=9cm]{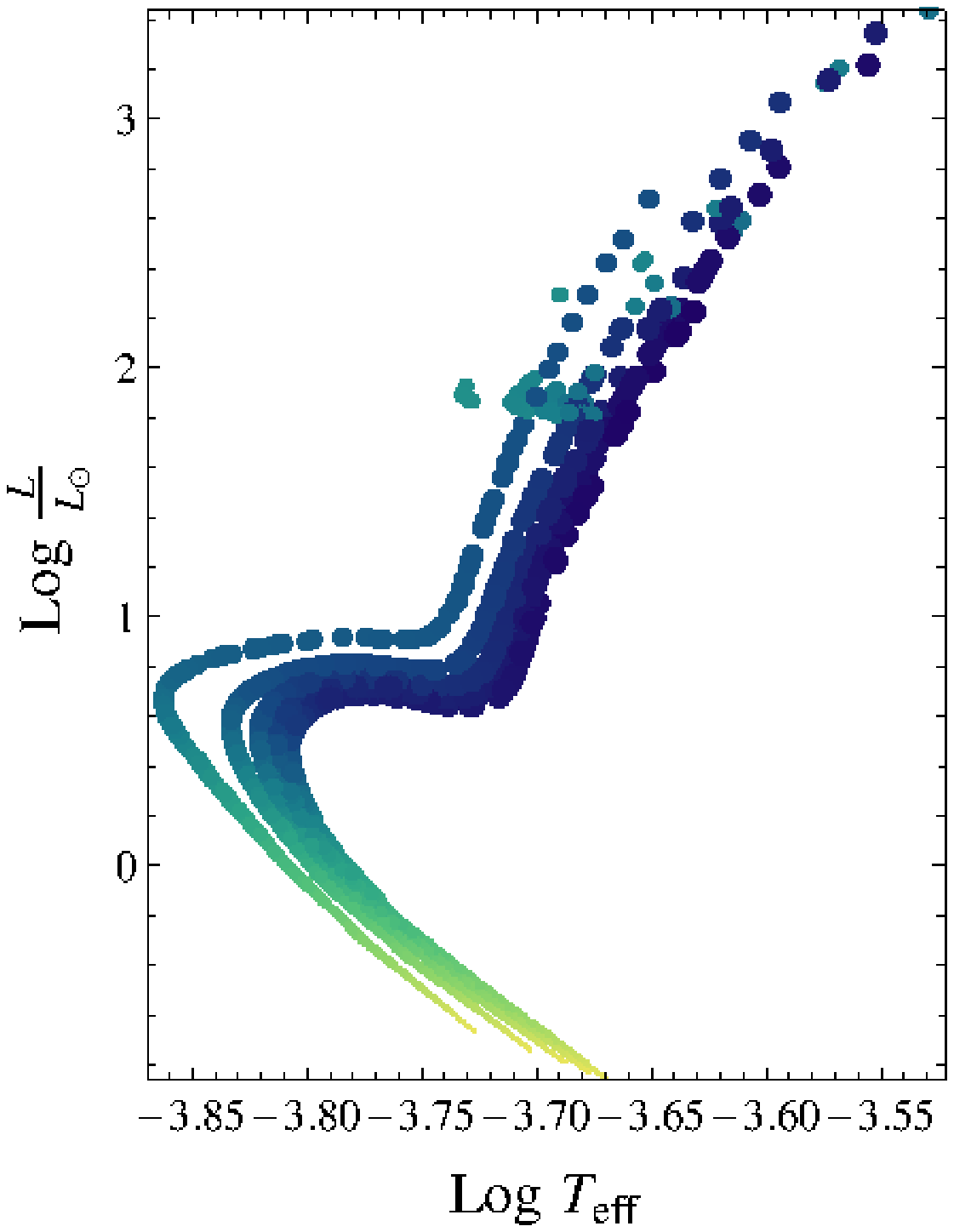} 
\includegraphics[width=8cm, height=10cm]{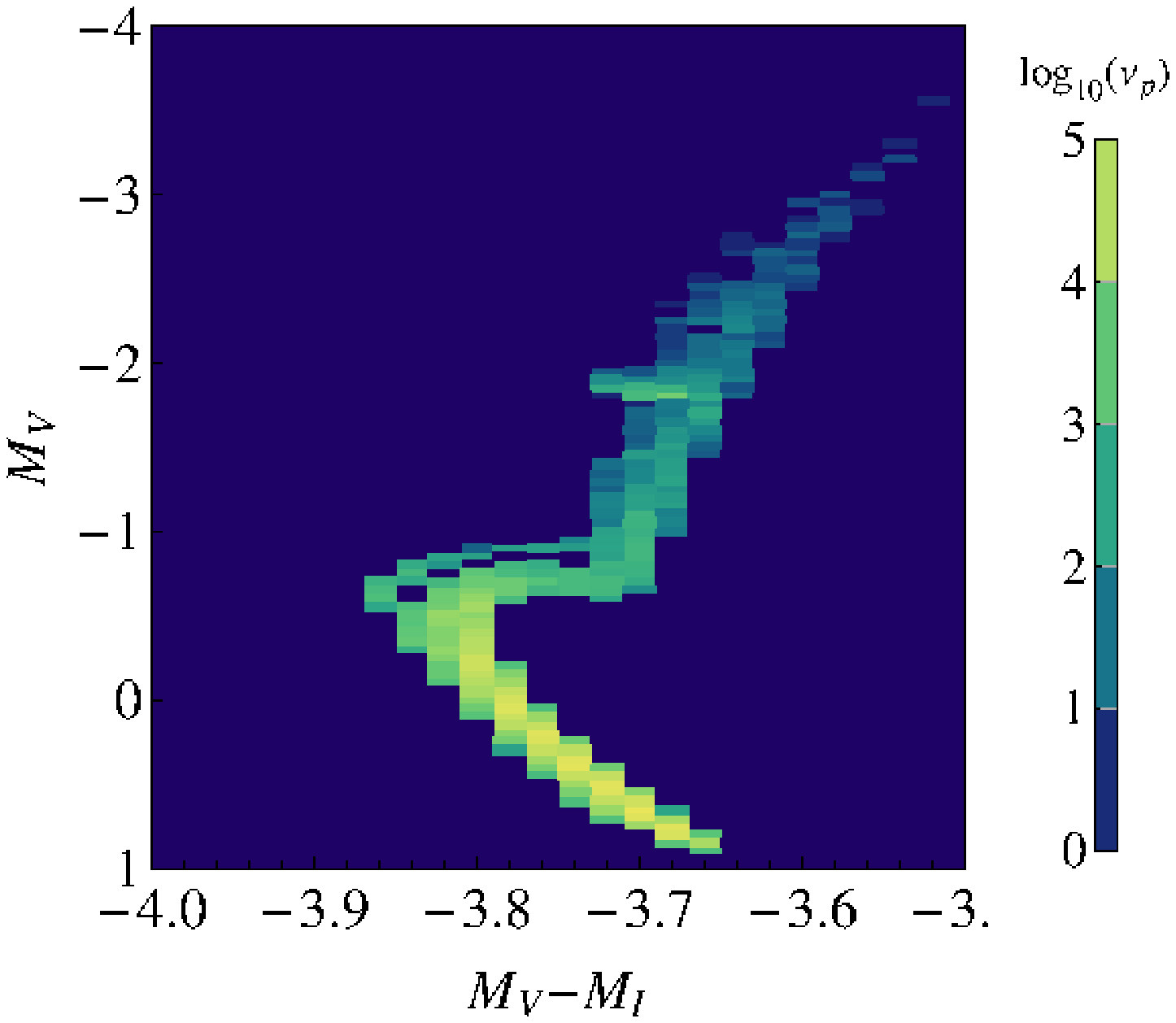}}
\caption{\textsf{Left panel}:  The HRD of SSPs with assigned age and metallicity. The size of the circles indicates the current star mass along the isochrones and/or SSP. The stellar models in use are taken from the library  of \citet{2008A&A...484..815B} and \citet{2009A&A...508..355B}. The colour code refers to the size of the dots: the smaller is the dot, the more yellow it is and the smaller the mass of the plotted stars. The bigger the dot, the bluer its colour and the bigger the mass of the stars. The mass considered in the models of this example ranges from 0.15 to 20 M$_\odot$.
\textsf{Right panel}: The same SSPs displayed in the left panel but in the observational CMD $M_V$ vs $M_V-M_I$. The CMD is subdivided to a discrete grid of elemental cells, in each of which a certain number of stars fall. The stars in each cell may belong to different SSPs. The size of the cell is large clearly show evidence  the ``out-of-focus'' effect intrinsic to this tessellation technique. Colour-code spans account for the number of stars in the cells from 0 (the bluer colour) to $10^5$ (the light yellow colour).}
\label{SSPBinned}
\end{figure*}

Finally we are left to specify the initial distribution of mass (IMF) and the adopted database of stellar models:

\begin{itemize}
	\item \textsf{The initial mass function}: To calculate an SSP, we need an IMF. The most popular  IMF is
given by
\begin{equation}\label{Eq18}
	\xi \left( M \right) = {\xi _0}\sum\limits_{i = 1}^k {{M^{ - {x_i}}}{\Delta _i}}
\end{equation}
with
\begin{equation}\label{Eq19}
	{\Delta _i} = \left\{ {\begin{array}{*{20}{c}}
	  1&{M \in \Delta {M_i}} \\
	  0&{M \notin \Delta {M_i}}
	\end{array}} \right.
\end{equation}
with ${\xi _0}$ normalisation factor of the IMF, $\Delta {M_i} = \left\{ {{M_i}|{M_i} \in \left[ {{M_{{\text{low}},i}},{M_{{\text{up}},i}}} \right]} \right\}$, with  ${M_{{\text{low}},i}}\left( {{M_{{\text{up}},i}}} \right)$ lower (upper)  limit for the ${i^{th}}$-mass interval and $\Delta {M_i} \cap \Delta {M_j} = \emptyset $ for $i \ne j$. The case $k = 1$ in Eq. \eqref{Eq18} is often referred to as Salpeter's IMF with power-law slope  ${x_i} = 2.35$ $\forall i$, the case $k = 3$ is referred as Kroupa's IMF with its specific slope and intervals \citep[see e.g.,][]{2001MNRAS.322..231K} or in general any IMF function can be approximated by a sum as in Eq. \eqref{Eq18} for a suitable choice of $k$. Therefore, in the general case, the total number of stars plotted in an SSP is
\begin{equation}\label{Eq20}
	\begin{gathered}
	  {N_{{\text{SSP}}}} = \int_{{M_{{\text{low}}}}}^{{M_{{\text{up}}}}} {{\xi _0}\sum\limits_{i = 1}^k {{\Delta _i}{M^{ - {x_i}}}} dM}  \\
	   = {\xi _0}\sum\limits_{i = 1}^k {{\Delta _i}\int_{{M_{{\text{low}}}}}^{{M_{{\text{up}}}}} {{M^{ - {x_i}}}dM} }  \\
	   = \begin{array}{*{20}{c}}
	  {{\xi _0}\sum\limits_{i = 1}^k {\frac{{{\Delta _i}}}{{{x_i} - 1}}\frac{{M_{{\text{up}}}^{{x_i} - 1} - M_{{\text{low}}}^{{x_i} - 1}}}{{{{\left( {{M_{{\text{low}}}}{M_{{\text{up}}}}} \right)}^{{x_i} - 1}}}}} }&{{x_i} \ne 1}
	\end{array} \\
	\end{gathered}
\end{equation}
or ${N_{{\text{SSP}}}} = {\xi _0}\sum\limits_{i = 1}^k {{\Delta _i}\log \left( {\frac{{{M_{{\text{up}}}}}}{{{M_{{\text{low}}}}}}} \right)} $ for ${x_i} = 1$. If we choose to sample all  star-phases of higher luminosity (e.g., asymptotic giant branch (AGB) and planetary nebula (PNs)) with ${N_{{\text{SSP}}}} \cong  {{{10}^5}}$ stars  it becomes immediately evident how the graphical realization of a CMD for  an N-body simulation encounters serious problems once the number of star-particles involved, N, is high, say ${N_{{\text{SSP}}}} \times \text{N} \cong O\left( {{{10}^{11 \div 13}}} \right)$ for ${\text{N}} \sim {\text{1}}{{\text{0}}^{6 \div 8}}$. \\ 

\item \textsf{The database of SSPs}: We briefly report here on the data base of SSPs that was calculated for the purposes of this study. The stellar models used are those by \citet{2008A&A...484..815B,2009A&A...508..355B}, which cover a wide grid of helium $Y$, metallicity $Z$, and enrichment ratio $\Delta Y / \Delta Z$. The associated isochrones include the effect of mass loss by stellar wind and the thermally pulsing AGB phase according to the models calculated by \citet{2007A&A...469..239M}.
The code used is the last version of YZVAR developed over the years by the Padova group used in many studies \citep[for instance][]{1981A&A....98..336C,1986sfdg.conf..449C,1986MmSAI..57..507C,1989A&A...219..167C,1995A&A...301..381B,1995A&A...295..655N,1996ApJ...469L..97A,2001AJ....121.1013B,
2003AJ....125..770B} and was recently extended to   obtain isochrones and
SSPs in a large region of the $Z-Y$ plane. The details on the interpolation scheme at a given $\Delta Y/ \Delta Z$ are given in \citet{2008A&A...484..815B,2009A&A...508..355B}.

The present isochrones and SSPs are in the Johnson-Cousins-Glass system as defined by \citet{1990PASP..102.1181B}  and \citet{1988PASP..100.1134B}.  The formalism adopted to derive the bolometric corrections  is described in \citet{2002A&A...391..195G}, while the definition and values of the zero-points are described in
\citet{2007A&A...469..239M} and \citet{2007A&A...468..657G} and will not be repeated here.

Suffice it to recall that the bolometric corrections  stand on an updated and extended library of stellar spectral fluxes. The core of the library now consists of the ODFNEW ATLAS9 spectral fluxes from \citet{2003IAUS..210P.A20C}, for ${T_{{\text{eff}}}} \in \left[ {3500,50000} \right]$ K, ${\log _{10}}g \in \left[ { - 2,5} \right]$ (with g the surface gravity), and scaled solar metallicities ${\text{[M/H]}} \in \left[ { - 2.5, + 0.5} \right]$. This library is extended at the intervals of high $T_{\text{eff}}$ with pure black-body spectra. For lower $T_{\text{eff}}$, the library is completed with the spectral fluxes for M, L and T dwarfs from \citet{2000ASPC..212..127A}, M giants from \citet{1994A&AS..105..311F}, and finally the C star spectra from \citet{2001A&A...371.1065L}. Details about the implementation of this library, and in particular about the C star spectra, are provided in \citet{2007A&A...469..239M}. It is also worth mentioning that in the isochrones we applied the bolometric corrections derived from this library without making any correction for the enhanced He content which has been proved by \citet{2007A&A...468..657G} to be low in most common cases.

The database of SSP  covers the space of existence $\mathbb{E}$ of a generic CPS. The number of ages ${N_\tau }$  of the SSPs are sampled according to a law of the type $\tau  = i \times {10^j}$ for $i = 1,...,9$ and $j = 7,...,9$, and for $N_Z$ metallicities are $Z = \left\{ {0.0001,0.0004,0.0040,0.0080,0.0200,0.0300,0.0400} \right\}$. The helium content associated to each choice of metallicity is according to the enrichment law $\Delta Y \Delta Z =2.5$.  Each SSP  was calculated allowing a small age range around the current value of age given by $\Delta\tau = 0.002\times 10^j $ with $j=7,...9$. In total, the data base  contains
 ${N_\tau } \times {N_Z} \cong 150$ SSP. This grid is fully sufficient to illustrate the method. For future practical application of it, finer grids of SSPs can be calculated and made available. Having done this, the  normalization constant ${\xi _0}$ of the SSPs remains defined.
Finally, for each SSP we computed the ``projected'' DF, i.e. the number of stars per elemental cell of the CMD,  using the value of  ${\xi _0}$.

Note that the method for generating the cumulative DFs of Eq. \eqref{nusomma} from SSPs does not depend on the particular choice for the data base of stellar tracks, isochrones, or photometric system.
Other libraries of stellar models and isochrones can be used to generate the database of SSP, the building blocks of our method. 

\end{itemize}

From the procedure outlined above, it is easy to understand how the use of the stellar DF is able to accelerate the construction of the CMD of a CSP. The reason is as follows: Instead of calculating an SSP made by $N_{{\text{SSP}}}$ stars for every star particle of the N-body simulation (i.e., for a total of N star particles) and counting the stars inside a given bin of magnitude and colour $\delta C\delta m$, i.e. summing over $\text{N} \times {N_{{\text{SSP}}}} \cong O\left( {{{10}^{11\div13 }}} \right)$ stars, with the above procedure, we need to calculate only ${N_\tau } \times {N_Z} \times {N_{{\text{SSP}}}} \cong O\left( {{{10}^7}} \right)$ stars. However, the number frequencies per cell of the CMD of our reference SSPs  are calculated once for all, whereas their combinations can be freely changed according to the underlying star formation history of the N-body simulation to investigate and the number of calculation required in this method is then just $O\left( {\text{N}} \right) \ll {10^{11 \div 13}}$.

The left panel of Fig.\ref{SSPBinned} shows a few SSPs for the solar metallicity $Z=0.02$ and helium content $Y=0.28$ in the theoretical Hertzsprung-Russell diagram (HRD). The size of the  dots is proportional to the stellar mass running along the SSPs. The same SSPs are translated into the  $M_V$ vs $M_V-M_I$ plane displayed in the right panel of Fig. \ref{SSPBinned} in which the cell tessellation is evident. The "out-of-focus" effect is due to grouping the stars of different SSPs into the same cell. No photometric errors are applied.

\subsection{Simulation of photometric errors and completeness}
We finally mention that real data on the magnitudes (and colours) of the stars are affected by photometric errors, whose amplitude in general increases at decreasing luminosities (increasing magnitude). The photometric errors come together with the data themselves provided they are suitably reduced and calibrated. Photometric errors can be easily simulated in our theoretical CMDs. Suppose we know the errors affecting our magnitudes
in the two pass-bands used to build the CMD, and that these are a function of the magnitude itself. Let us indicate the errors by $\delta m_\alpha(m)$, and $\delta m_\beta(m)$, with $\alpha$ and $\beta$ the two pass-bands. When plotting each point of the SSPs on the observational CMDs, the magnitudes are changed by the quantity representing the errors, e.g.:
\begin{eqnarray}
      m'_\alpha(m)&=& m_\alpha(m) - \left(\frac{1}{2}-r\right)\delta m_\alpha(m) \nonumber \\
      m'_\beta(m)&=& m_\beta(m) - \left(\frac{1}{2}-s\right)\delta m_\beta(m),
\end{eqnarray}
where $r$ and $s$ are two randomly drawn numbers (e.g., from uniform or Gaussian distribution) comprised between 0 and 1. The star frequencies per elemental cell of the CMD are calculated after applying the correction for photometric errors to the reference SSPs.

To compare the real observational data with the theoretical model we have to know the completeness of the former as a function of the magnitudes and pass-band  \citep{1988AJ.....96..909S,1995AJ....110.2105A}. This is a long known problem that does not require any particular discussion in the context of this paper, and tabulations of the completeness factors must be supplied in advance together with the correction for photometric errors. We mention here is that correcting for completeness will alter the DF of stars in the cells of the observational CMD we aim to analyse \citep[e.g.,][]{2011A&A...530A..58C}. These completeness factors must be supplied by the user of our method in connection with the specified problem.

\begin{figure*}
\centering{
\includegraphics[width=8cm, height=9cm]{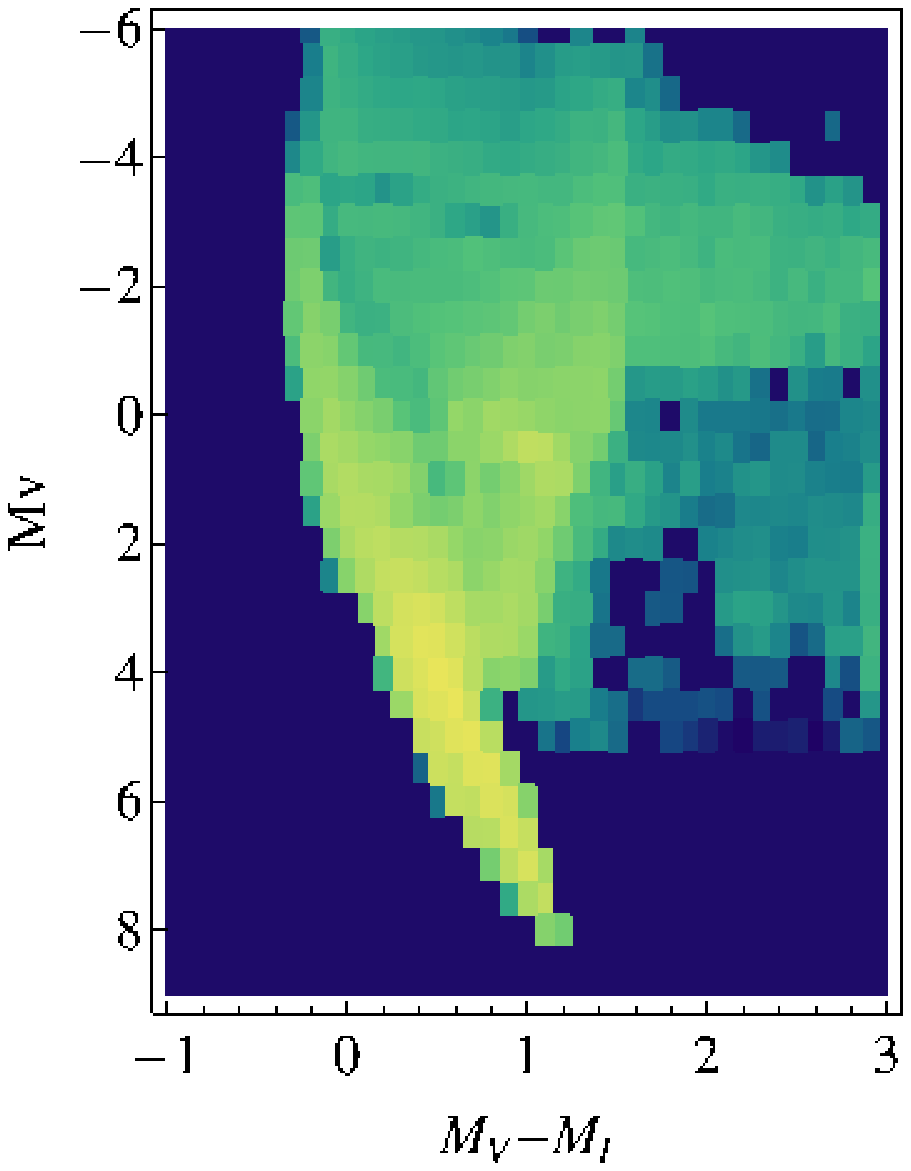}
\includegraphics[width=9cm, height=6.5cm]{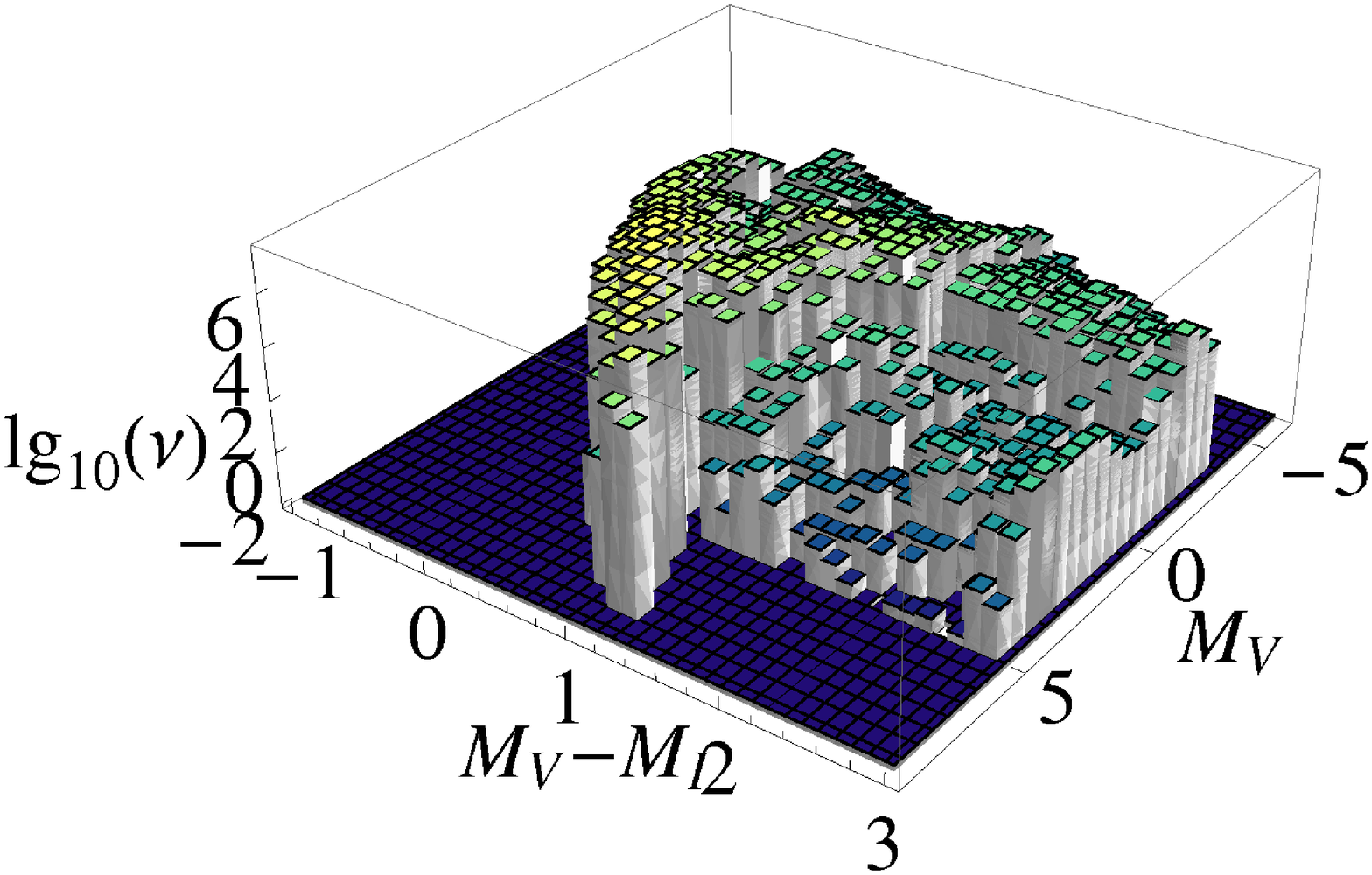}}
\caption{\textsf{Left panel}: The CSP for an N-body simulation of a disc galaxy. Colour code refers to the frequency of star per bin shown in the companion right panel.  \textsf{Right panel}: The histogram of DF for the CSP shown in the left panel. In this figure, one immediately  captures the concept of DF described in the text.  The characteristic peaks  corresponding to MS and RGB stars are shown in yellow. The CMDs have the coordinates of the absolute magnitudes $M_V$ and the colours $M_V-M_I$.}
\label{Csp_Histo}
\end{figure*}

\begin{figure*}
\resizebox{\hsize}{!}{\includegraphics{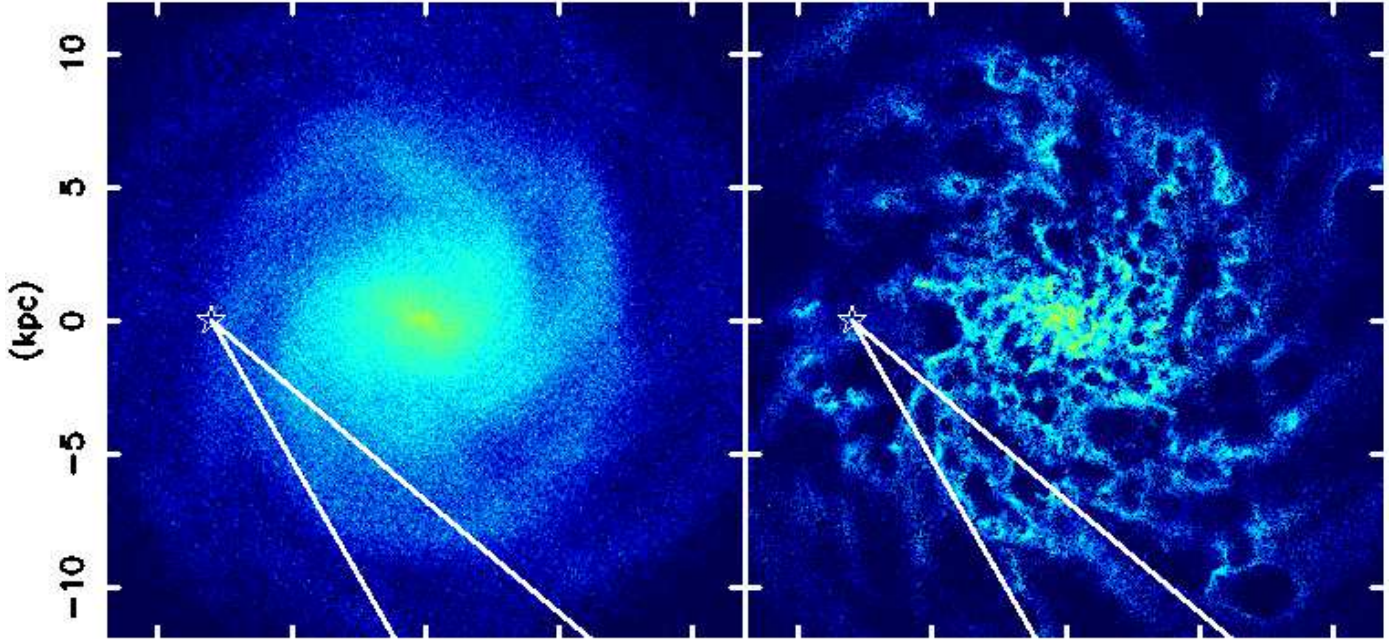}}
\caption{
Snapshot of the simulated disc galaxy whose face-on views of the stellar and gas discs are shown in the left and right panels, respectively. Our assumed location of the Sun is highlighted by a star symbol. The white lines indicate the selected region for constructing the CMD shown in Fig. \ref{Gaia}. Colour code ranges from blue (the lowest density) to red (the highest density).
}
\label{snapshot}
\end{figure*}

\section{Generating a CMD with tessellation}

To explain  the concept of CMD tessellation, we simulated a CMD using the CSP of NB-TSPH simulation (to be described in more detail below). The analysed field contains $8.8 \times {10^4}$ star-particles of the same mass and known age and metallicity (to each of which an SSP can be associated with the same mass, age and metallicity, see Section \ref{Sp4Nb}). For the purpose of this example we show only the absolute luminosity and magnitude, i.e. no distance dependence. The results of applying Eq. \eqref{nusomma} are shown in Fig. \ref{Csp_Histo}. The left panel displays the frequency distribution of CSP in the $(M_V-M_I)$ vs $M_V$ CMD, while the right panel shows the histogram. 
 The regions occupied by stars in the main sequence are clearly evident: Long-lived phases display a higher number of stars, on the other hand, red giant, red clump, and asymptotic giant phases are seen in low effective temperatures (red colours) and shows lower frequency due to their short lifetime. These features are similar to a typical observed CMD of the stellar populations in nearby galaxies, e.g. the Magellanic Clouds, M31 and others, where for all stars in a given galaxy the distance is nearly constant.
%

\section{Application to N-body simulations}\label{NB_simul}

To demonstrate the method, we applied it to an N-body simulation. The simulation was carried out with an updated version of our original NB-SPH code, GCD+ \citep{2003MNRAS.340..908K,2012MNRAS.tmp.2756R}. We initially set up an isolated Milky Way-sized disc galaxy that consists of gas and stellar discs with no bulge component in a static dark matter halo potential, following \citet{2012MNRAS.tmp.2756R}. Note that this simulation was used for demonstration purpose, and was not meant to reproduce the Milky Way. We used the standard Navarro-Frenk-White dark matter halo profile \citep{1997ApJ...490..493N} with the total mass of $M_{\rm tot}=1.5\times10^{12}$ M$_{\odot}$ and the concentration parameter of $c=12$. The mass, scale length and scale high of the stellar disc were assumed to be $M_{\rm d,s}=4.0\times 10^{10}$ M$_{\odot}$, $R_{\rm d,s}=2.5$ kpc and $z_{\rm d,s}=350$ pc. The mass and scale length of the gas disc was $M_{\rm d,g}=1.0\times 10^{10}$ M$_{\odot}$, $R_{\rm d,g}=4.0$ kpc.
We initially set 400,000 particles to the gas disc and 1,600,000 particles to the stellar disc.
Therefore, our baryon particle mass was $M_{\rm p}=2.5\times 10^{4}$ M$_{\odot}$. We applied the threshold density of $n_{\rm H}=1.0$ cm$^{-3}$ for star formation. \citet{2012MNRAS.tmp.2756R} demonstrated that the star formation in a disc is quite sensitive to the parameters of star formation efficiency, $C_*$, energy for supernova, $E_{\rm SN}$, and stellar wind feedback energy, $E_{\rm SW}$.
We here applied $C_*=0.1$, $E_{\rm SN}=10^{51}$ erg and $E_{\rm SW}=10^{37}$ erg s$^{-1}$.
Fig. \ref{snapshot} shows a snapshot of the simulation. Strong feedback creates many bubbles in the gas disc.

\begin{figure}
\resizebox{\hsize}{!}{\includegraphics{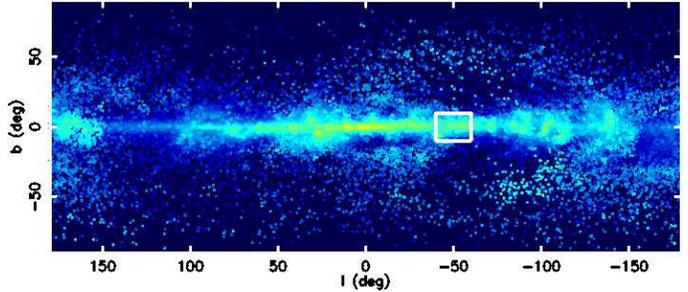}}
\caption{
Gas distribution in a Galactic coordinate of the simulated galaxy shown in Fig. \ref{snapshot}. The region enclosed by the white lines are the selected region for constructing the CMD shown in Fig. \ref{Gaia}. The colour coding is by density, so that bluer and more yellow points represent higher and lower density particles respectively. 
}
\label{snapshotcut}
\end{figure}

The simulated galaxy shows a small bar and several spiral arms. We set the Sun at $(x,y)=(-8,0)$ kpc (star symbol in Fig. \ref{snapshot}), and selected star-particles in the region of the simulated galaxy ``equivalent longitude'' $300<l<320$ and ``latitude'' $-10<b<10$ (with implicit reference to the Milky Way galaxy coordinates system ${l,b}$), which is enclosed by white lines in Figs. \ref{snapshot} and \ref{snapshotcut}.
For each star-particle we measured the distance and extinction from the simulation. To measure the extinction, first the column density for each star-particle was calculated by summing up the line of sight column density of all gas particles between the star-particle and the position of the Sun, using the SPH weighting scheme \citep{2007ApJ...663...38K}. We then converted the column density to the extinction, using $N_{\rm H}=1.9\times10^{21} {\rm cm}^{-2} \times A_V ({\rm magnitude})$.

\begin{figure}
\resizebox{\hsize}{!}{\includegraphics{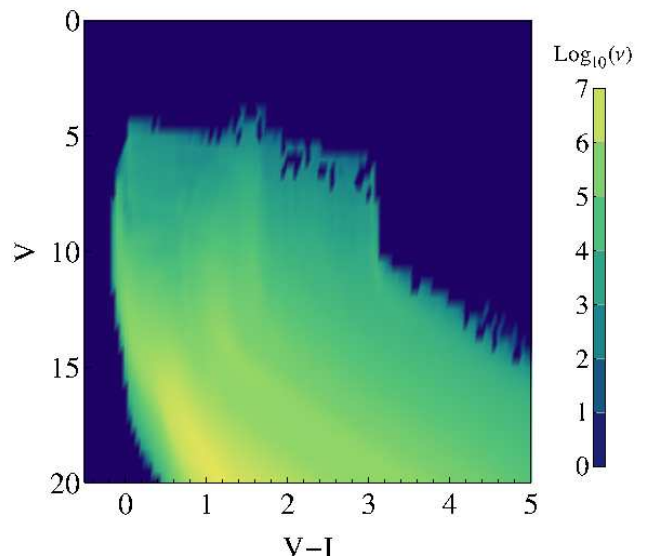}}
\caption{Observational CMD in apparent magnitudes and colours. The colour-coding is by density, so that bluer and more yellow points represent higher and lower density particles, respectively.}
\label{Gaia}
\end{figure}

The simulation of the CMD for  the selected star-particles is shown in Fig. \ref{Csp_Histo} where with a magnitude cut for true stars at above $V>20$ mag  about $8.8 \times 10^4$ star-particles within the galaxy coordinates defined above are retained. We recall that Fig. \ref{Csp_Histo} shows absolute magnitude and luminosity, which are independent of the distance. To build the observational CMD we need to take into account distance and extinction for each star-particle, which are measured for each particle as described above.  The resulting ``observational'' CMD   is a shown in Fig. \ref{Gaia}. With the IMF we assumed that the resulting CMD is representative of about $1.08 \times 10^{10}$ true stars. The difference in distance and extinction for each star-particle leads to a more smoothly distributed CMD. As a result, the groups of main sequence  stars and red giant stars are hardly recognizable in Fig. \ref{Gaia}. This is often observed in the photometric data of Galactic stars.

\section{Concluding remarks}\label{ConcDisc}
We first extended the theory of the stellar populations
to include the phase-space description. We described
the concepts of composite stellar population, simple
stellar population, star formation rate, initial mass function,
etc. using the language of Statistical Mechanics. Now it is
fair to ask what we have gained from this rather formal
approach. These techniques have proven to be extremely
powerful in other branches of physics e.g., for the
study of relaxation phenomena, electrical conduction, irreversible
processes \citep[see e.g.,][]{LL} or in general for the development of
the thermodynamics of non-equilibrium system \citep[see e.g.,][]{dGM} while have been more limited in the treatment of long-range forces e.g., the statistical mechanics of gravitational systems \citep[see, e.g.,][]{1968MNRAS.138..495L, Katz:astro-ph0212295}, therefore it is important
to understand to what extent they can be applied to the stellar populations. 
The mutual benefit between Statistical Mechanics and stellar populations theory consists of the temporal evolution of the existence space we introduced,  the projection of which onto the CMD is governed by the fuel consumption theorem that drives the relative number of stars in different evolutionary stages, hence cells of the CMD.
Moreover, studying the   coupling between dynamics and theory of the stellar populations  imprinted  in  the existence space, may reveal  unexplored connections.

The second achievement of this paper is indeed the application
of this theory to develop a method for handling
the synthetic CMDs that one would generate from N-body simulations,
which are to be compared with the CMDs of
real galaxies when they are resolvable into stars. The method based
on the concept of star frequencies per elemental cell of the
CMD can easily simulate observational CMDs that contain
huge numbers of stars. This study extends and completes
other similar studies in literature that dealt with synthetic
CMDs such as those presented by in Carraro et al. (2001);
\citet{2012A&A...542A..17P}.
This technique aims to interface N-body simulations to photometric
observations and is developed with the perspective of the
ever improving capabilities of the N-body simulations in the future.
Finally, thanks to  its agility, the CMD tessellation method can be suitably interfaced with galaxy models based on star counts  \citep[see for instance ][]{2006A&A...451..125V} and other  hybrid techniques based on the different approaches already present in literature such as the Galaxia model \citep[see e.g., ][]{2011ApJ...730....3S} or the Besancon model  \citep[see e.g. ][]{2003A&A...409..523R}  and it represents a natural extension of previous work  \cite[e.g. ][]{2010A&A...518A..43T}.

The tessellation code for generating the CMD for  N-body simulations  is available upon request from the authors.

\begin{acknowledgements}
SP acknowledges D. Crnojevi\'c and J. Hunt for careful reading of an early version of this manuscript. The authors thank the anonymous referee for the constructive report. 
\end{acknowledgements}

\bibliographystyle{aa} 
\bibliography{BiblioArt} 
\end{document}